\documentclass[12pt,preprint]{aastex}

\def\ltsima{$\; \buildrel < \over \sim \;$}
\def\gtsima{$\; \buildrel > \over \sim \;$}
\def\lsim{\lower.5ex\hbox{\ltsima}}
\def\gsim{\lower.5ex\hbox{\gtsima}}
\def\lapp{\ifmmode\stackrel{<}{_{\sim}}\else$\stackrel{<}{_{\sim}}$\fi}
\def\gapp{\ifmmode\stackrel{>}{_{\sim}}\else$\stackrel{<}{_{\sim}}$\fi}

\def\teff{\ensuremath{T_{\rm eff}}}

\def\teff{\ensuremath{T_{\rm eff}}}

\def\recio{\ensuremath{{\rm Log}(T_{\rm eff_{HB}})}}
\def\dotter{$\Delta (V-I)$}
\def\feh{{\rm [Fe/H]}}
\def\gyr{{\rm Gyr}}

\newbox\grsign \setbox\grsign=\hbox{$>$} \newdimen\grdimen \grdimen=\ht\grsign
\newbox\simlessbox \newbox\simgreatbox
\setbox\simgreatbox=\hbox{\raise.5ex\hbox{$>$}\llap
     {\lower.5ex\hbox{$\sim$}}}\ht1=\grdimen\dp1=0pt
\setbox\simlessbox=\hbox{\raise.5ex\hbox{$<$}\llap
     {\lower.5ex\hbox{$\sim$}}}\ht2=\grdimen\dp2=0pt
\def\simgreater{\mathrel{\copy\simgreatbox}}
\def\simless{\mathrel{\copy\simlessbox}}
\newbox\simppropto
\setbox\simppropto=\hbox{\raise.5ex\hbox{$\sim$}\llap
     {\lower.5ex\hbox{$\propto$}}}\ht2=\grdimen\dp2=0pt

\shorttitle{GALEX integrated colors}
\shortauthors{Dalessandro et al.}
 
\begin{document} 
\title{UV Properties of Galactic Globular Clusters with GALEX II. Integrated colors.\\
}
\author{
Emanuele Dalessandro\altaffilmark{1},
Ricardo P. Schiavon\altaffilmark{2},
Robert T. Rood\altaffilmark{3},
Francesco R. Ferraro\altaffilmark{1},
Sangmo T. Sohn \altaffilmark{4},
Barbara Lanzoni \altaffilmark{1},
Robert W. O'Connell \altaffilmark{3}}
\affil{\altaffilmark{1} Dipartimento di Astronomia, Universit\`a degli Studi di Bologna, via
	Ranzani 1, I-40127, Bologna, Italy}
\affil{\altaffilmark{2} Gemini Observatory, 670 N. A'Ohoku Place, Hilo, HI 96720, USA}	
\affil{\altaffilmark{3} Astronomy Department, University of Virginia, P.O. Box 400325,
	Charlottesville, VA 22904, USA }  
\affil{\altaffilmark{4} Space Telescope Science Institute, 3700 San Martin Dr., Baltimore, MD
	21218, USA}

\date{23 August, 2012}

\begin{abstract}
We present ultraviolet (UV) integrated colors of 44 Galactic globular
clusters (GGCs) observed with the Galaxy Evolution Explorer (GALEX) in
both $FUV$ and $NUV$ bands.  This data-base is the largest homogeneous
catalog of UV colors ever published for stellar systems in our Galaxy.
The proximity of GGCs makes it possible to resolve many individual
stars even with the somewhat low spatial resolution of GALEX.  This
allows us to determine how the integrated UV colors are driven by hot
stellar populations, primarily horizontal branch stars and their
progeny.  The UV colors are found to be correlated with various
parameters commonly used to define the horizontal branch morphology.
We also investigate how the UV colors vary with parameters like
metallicity, age, helium abundance and concentration. We find for 
the first time that 
GCs associated with the Sagittarius dwarf galaxy have $(FUV-V)$ colors 
systematically redder than GGCs with the same metallicity. 
Finally, we speculate about the presence of an interesting trend,
suggesting that the UV color of GCs may be correlated with 
the mass of the host galaxy, in the sense that more massive 
galaxies possess bluer clusters.
\end{abstract} 


\section{INTRODUCTION}
The main contributors to the UV emission from any stellar system are
the hottest stars. Indeed blue horizontal branch (HB) stars are well
known to be among the hottest stellar populations in globular clusters
(GCs) and contribute substantially to the UV radiation observed from
old stellar systems (Welch \& Code, 1972).  
Later on, two other sub-classes of post-HB stars were found to be important 
contributors to far-UV
($FUV$, $\lambda \sim 1500$--1600\,\AA) radiation (e.g., Greggio and Renzini 1990;
Dorman et al. 1995, hereafter DOR95; Han et al. 2007).
The hottest HB
stars (extreme HB, EHB) have such a small envelope mass that most of
their post-He-core burning phase takes place at high effective
temperature (\teff), during the so called "AGB-manqu\'e phase", and
these stars never return to the asymptotic giant branch (AGB).
Another group of UV-bright stars is that of post-early AGB stars,
which after a brief return to the AGB, spend the bulk of their helium
shell burning phase at high \teff. In systems with only red HB a small floor level
of $FUV$ is provided by post-AGB stars, which  
evolve to the AGB phase with an higher envelope mass where they undergo 
thermal pulses and eventually lose their envelopes moving at higher temperatures 
at constant luminosity.
The relative contributions of the various types of stars and
the factors that might lead to larger or smaller populations of
UV-bright stars have remained an open question (Greggio and Renzini 1990;
DOR95; Lee et al.  2002; Rich et al. 2005; Sohn et al. 2006).

In distant extragalactic systems one can ordinarily observe only the
integrated light of unresolved stellar populations, from which the
hope is to gain knowledge about the underlying stellar population. Galactic
globular clusters (GGCs)
play an important role in understanding the integrated UV colors of
extragalactic systems, especially the so called "UV-upturn" observed in the spectral energy 
distributions of elliptical galaxies (Code \& Welch 1979; de Boer 1982; Bertola et al. 1982;
Greggio \& Renzini 1990; O'Connell 1999).
First of all,
GCs are the closest example in nature to a single stellar
population (SSP): a system of coeval stars with similar chemical
composition\footnote{Although there is now a general consensus that
  the formation of GGCs may have been more complex than previously
  thought (based for example, on the detection of chemical
  inhomogenites in light elements, Carretta et al. 2009a, and the
  existence of multiple populations, Piotto 2009), their stellar content
  has been found to be quite homogeneous in terms of iron
  abundance. Indeed only two GC-like stellar systems have been found
  to host multiple populations with significant ($>0.5$ dex) spread in
  the iron abundance and (possibly) age: $\omega$ Centauri (Norris et
  al. 1996, Lee et al. 1999, Ferraro et al. 2004) and Terzan~5 (Ferraro
  et al. 2009).}.  Moreover GGCs span a large range of metallicities,
a small range of ages, and perhaps some range of helium
abundance. Hence they can be used to test the stellar evolution theory, which 
in turn is one of the basic ingredients
of the models used to interpret the integrated light of distant galaxies.
GGCs are relatively nearby objects (more than
$\sim90\%$ are located at distances $r<30$\,kpc), so their populations
can be easily resolved.  With typically more than 100,000 stars, even
relatively short-lived evolutionary stages are sampled.  We can
directly observe the properties of individual stars and measure the
population ratios for objects in different evolutionary stages. In
particular we can study the impact of hot and bright populations (as
the AGB-manqu\'e stars) on the integrated UV light of GGCs and then
use them as crucial local templates for comparison with integrated
properties of distant extragalactic systems. In fact comparing
features in the color-magnitude diagrams (CMDs) of well known and
resolved GGCs with integrated quantities can lend important {\it model
  independent} insights into the nature of extragalactic systems.

Integrated UV photometry of GGCs has previously been obtained by the
\emph{Orbiting Astronomical Observatory} (OAO~2; Welch \& Code 1980),
\emph{Astronomical Netherlands Satellite} (ANS; van Albada, de Boer \&
Dickens 1981), \emph{Ultraviolet Imaging Telescope} (UIT; Hill et
al. 1992; Landsman et al. 1992; Parise et al. 1994; Whitney et
al. 1994), and \emph{International Ultraviolet Explorer} (IUE;
Castellani \& Cassatella 1987). Using a large, but heterogeneous,
collection of data obtained with some of these telescopes along with
population synthesis models, DOR95 showed how the UV colors varied
with parameters like metallicity and how they compare with elliptical
galaxies. They showed how the UV colors of GGCs could plausibly be
produced by hot HB stars and their progeny.

At the time of DOR95, UV photometry of individual stars in GGCs was
available for only a few clusters. That situation has changed
dramatically.  Our group alone has already published HST UV photometry
for a dozen of GGCs (see Ferraro et al.  1997, Ferraro et al. 1998,
Ferraro et al.  1999, Ferraro et al.  2001, Ferraro et al. 2003,
Lanzoni et al. 2007, Dalessandro et al.  2008, Rood et al. 2008) and
obtained data for an additional 32 clusters in HST Cycle 16S (GO11975,
PI: Ferraro; see Contreras et al. 2012, Sanna et al. 2012 and
references therein).

More recently, we have secured observations of 44 GGCs during three
observing cycles with the \emph{Galaxy Evolution Explorer} (GALEX).
This is the largest homogeneous sample ever collected for GGCs in
UV so far.  In Schiavon et al. (2012; hereafter Paper I) we presented
photometry and CMDs for these clusters.  Here
we present integrated UV magnitudes and colors (\S~2) for each
cluster, and we describe (\S~3) how they are
affected by HB class, metallicity, age, and possibly structural
parameters (mass, density, central relaxation time, etc.). In \S~4
we compare our data with observations of GCs in M31
and M87.

\section{OBSERVATIONS and DATA ANALYSIS}
Images for 38 GGCs were obtained as part of GI1 and GI4 GALEX programs
(P.I. R. P. Schiavon) and supplemented with somewhat shallower
exposures for 8 GGCs obtained as part of program GI3 (P.I.,
S. Sohn). Two clusters (NGC~6229 and NGC~6864) are in common between
the two programs.  A number of clusters that would have been very
interesting targets (including M13, M80, $\omega$\,Centauri, NGC~6388,
and NGC~6441) could not be observed because of restrictions on the UV
background brightness, due to detector-safety considerations, which
prevented us from targeting at low Galactic latitudes, or near
positions of very bright UV sources.  With only one exception
(NGC~6273), images were obtained in both the $FUV$ ($\lambda_c =
1516$\,\AA) and near-UV ($NUV$, $\lambda_c = 2267$\,\AA) bands.  Thanks
to the wide field of view of GALEX ($\sim 1.2\deg$), it has been
possible to sample the full radial extent of most of the clusters.  A
few GCs, like NGC~104 (47Tuc) and NGC~5272 (M3), have tidal radii
larger than the GALEX field of view (McLauglhin \& van der Marel 2005,
hereafter MVM05), but these missing data do not significantly affect
our analysis.  Each GALEX image has been pre-processed by the standard
pipeline described in detail by Morrissey et al. (2005).  The raw
images (or ``count images'') have been converted to flux-calibrated
intensity images by applying relative response and flat-field
corrections and by scaling the flux by the effective area and exposure
time.

\section{INTEGRATED MAGNITUDES}
The integrated magnitudes have been obtained using two different
approaches: {\it (i)} by fitting the surface brightness profiles
(SBPs), and {\it (ii)} from direct aperture photometry (AP)
measurement on the images. In both approaches, the integrated
photometry will include the effects of objects below the detection 
threshold for individual stars. Comparing the results obtained with these
two different approaches we can correct for systematic effects or
biases.  In particular, AP can be affected by small number statistics
when a few very bright stars dominate the flux, especially at $FUV$
wavelengths (as for example, the case of the bright star in the very
central regions of 47~Tuc reported by O'Connell et al. 1997).  These
very bright objects may be cluster members (post-HB or post-AGB star)
or just foreground stars.  On the other hand, while colors obtained by
the SBP method are less affected by star count fluctuations, they may
suffer from uncertainties arising from the quality of the fit or the
determination of the center of the cluster. For these reasons we have
calculated integrated magnitudes with both the techniques, and we have
compared the results in detail.

In both cases, it was first necessary to determine the cluster
centers.  Since previous determinations (e.g., Harris 1996, 2010
revision -- hereafter H10; Noyola \& Gebhardt 2006; Goldsbury et al. 2010) either were
obtained with quite heterogeneous methods, or only partially overlap
our sample, we preferred to perform new estimates.  The plate scale of
GALEX ($1.5\arcsec\,{\rm pixel}^{-1}$ ) and the ``noise'' due to the
relatively small number of stars detected at UV band introduce
considerable uncertainty in the measurement of the barycenter of
resolved stars (e.g., Lanzoni et al. 2007).  We therefore determined
the center of the light distribution on $NUV$ images by following the same approach
described in Bellazzini (2007). For any given GC we picked an aperture
radius that, based on the size of cluster central region and its
density, was large enough to include a good fraction of the brightest sources.
We then
determined the light density (i.e. the fraction of observed flux per
unit area) in that aperture and iteratively varied the centering
position until maximum light density was found. 
We do not list the centers of light, since they have been obtained in
instrumental coordinates and therefore they could be used only when the images they 
refer to are
available. Moreover the poor spatial resolution of the GALEX detectors would 
have prevented 
to obtain a sufficient level of accuracy to make these values useful to the community 
for other applications.

\subsection{Integrated magnitudes from surface brightness profile fitting}
SBPs were obtained by using concentric annuli centered on the cluster
centers determined as described above. The number and the size of the
annuli were chosen for each cluster so as they sampled almost the same light fraction.
The flux calculation was performed by using standard IRAF tasks ({\it PHOT}, in the
DAOPHOT package, Stetson 1987). The flux calculated in each annulus
was normalized to the sampled area obtained adopting as pixel size
$1.5\arcsec$, as reported by Morrissey et al. (2005) for intensity map
images. The instrumental surface brightness profiles were then
transformed to the ABMAG GALEX magnitude system by applying the
zero-points reported in the GALEX on-line user's
manual\footnote{http://galexgi.gsfc.nasa.gov/docs/galex/instrument.html}
or in Section 4.2 of Morrissey et al. (2005).  As an example, the
observed SBP of NGC~6341 is shown in Figure~1 for both the $FUV$ and
$NUV$ channels. In this figure, solid squares are the annular surface
brightness measurements at each cluster radius, whereas open circles
represent the same values after performing sky subtraction as
described below.

\begin{figure}[!hp]
\begin{center}
\includegraphics[scale=0.8]{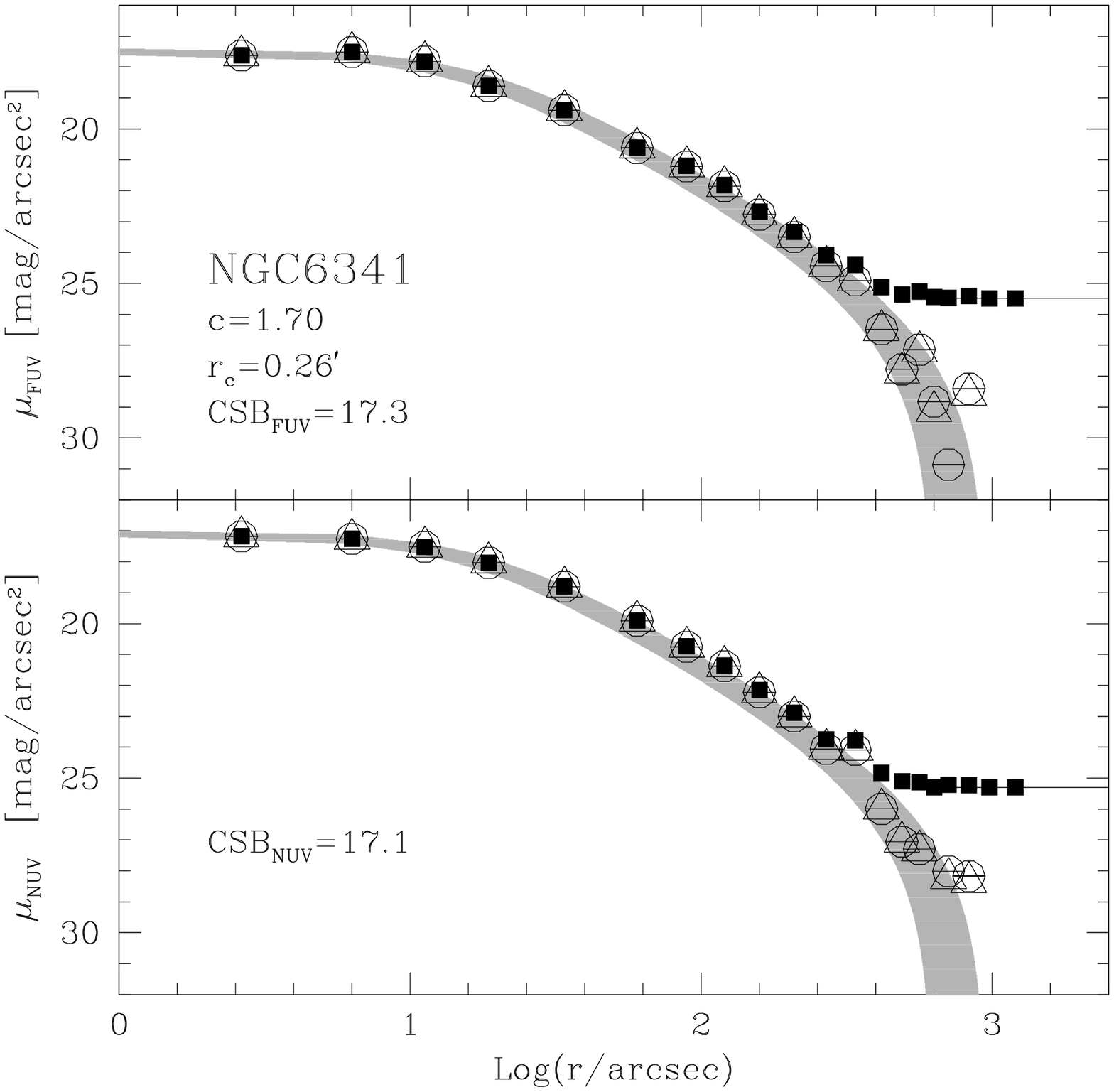}
\caption{$FUV$ and $NUV$ surface brightness profiles for NGC~6341. The
  black squares represent the observed raw profile, while open circles
  are the sky-subtracted values.  The horizontal black lines mark the
  estimated background level.  The background-subtracted profiles
  obtained by using standard IRAF routines are shown with open
  triangles, for comparison (see Section 3.1 for details).  The
  profiles are well reproduced by the King model (the grey region is defined by the structural 
  parameters uncertainties) with
  concentration parameter and core radius quoted by MVM05 and with
  central surface brightness (CSB) value providing the minimum
  $\chi^2$ (see labels).}
\label{figprof1}
\end{center}
\end{figure}

Using SBPs to obtain integrated colors requires the subtraction of the
background contribution, which is the combination of the sky
background emission (due to unresolved objects) and the resolved field
(back and foreground) sources. We determined the background as the
average value in the annuli at large radii where the observed SBPs show
a ``plateau'' (for $\log\,r>2.6$ in NGC~6341, see
Figure~\ref{figprof1}). The average has been obtained applying a sigma-clipping 
rejection to take into account
possible surface brightness fluctuations, since at low surface
brightness levels even a few 
background stars may cause strong deviations from the general behavior,
even in extended areas. We then subtracted the background light density from the
observed surface brightnesses in each radial bin thus deriving
``decontaminated'' SBPs.  As expected, background subtraction does not
significantly affect the central regions, but it substantially changes
the shape of the profile in the external parts.

To further check the robustness of our method, we compared our background
subtracted SBPs with those computed with standard IRAF packages ({\it
  PHOT} and {\it FITSKY}).  The open triangles in the upper panel of
Figure~\ref{figprof1} are obtained with background subtraction and
disabling the sigma-clipping routines.  They nicely overlap with the
open circles thus demonstrating that our approach is essentially
equivalent to the automatic procedure. Still we preferred to
independently fit the observed ``plateau'' in order to have a more
direct control of the background level which can be strongly variable
case by case. The errors for each bin are defined as the standard
deviations of the sampled fluxes.  We assumed that counts from both
the $NUV$ and $FUV$ detectors have a Poisson distribution. 

To determine the value of the central surface brightness, we have
fitted the $FUV$ and $NUV$ SBPs with single-mass King models over the
entire cluster extension.
For each system and each band we adopted
the core radius and concentration (together with their uncertainties)
quoted by MVM05, and we performed a mono-parametric fit, varying only
the central surface brightness: the value yielding the minimum
$\chi^2$ value was adopted as best-fit solution. Magnitudes were then
calculated by integrating the best-fit King model.  The extinction
coefficients used to correct the $FUV$ and $NUV$ fluxes are taken from
Cardelli et al.  (1989) and the $E(B-V)$ values were adopted from
H10.

In general the observed SBPs are very nicely reproduced by King models
with the adopted structural parameters (see the case of NGC 6341 in
Figure~\ref{figprof1}). Exceptions to this general trend are the post
core collapse clusters in our sample (NGC~6284, NGC~6342, NGC~6397 and
NGC~7099).  For these systems, which are not studied by
MVM05\footnote{Also Terzan~8 has not been analyzed by MVM05. In this
  case we used structural parameters quoted in H10.}, it is not
possible to fit in a satisfactory fashion the overall shape of the SBPs with
structural parameters typically adopted for post-core collapse
clusters ($c=2.50$ and $r_c=3-4\arcsec$; see H10 for example).
In these cases integrated magnitudes have
been obtained with aperture photometry only (see Section 2.2).
In addition we were not able to obtain a reliable measure in one or both
filters for Palomar~11 and Palomar~12, because of the low signal to
noise ratio of the images. We tested also the impact of using 
Wilson models (Wilson 1975) with structural parameters from MVM05. We find that
these models may give up to $0.4 - 0.5$ mag of difference for integrated 
magnitudes, but they have a null effect on colors.

The errors in the integrated magnitudes were obtained by propagating
the uncertainties affecting the structural parameters (as reported by
MVM05) and the central surface brightness determination. 
Another important source of uncertainty is that affecting the
determination of the cluster centers.  This can cause variations in
central surface brightness estimates that are independent of the other
parameters involved in the SBP fit.  To quantify the possible errors
due to mis-positioning of the cluster centers, we let the centers
iteratively vary by up to 5\arcsec\ from our determinations.  We found
maximum variations of $\Delta {\rm mag} \sim 0.1$ in the
$(FUV-NUV)$ color, mainly due to the errors in $FUV$ magnitudes.  
The colors derived are listed in Table~1.  Integrated
magnitudes in the $V$ band were also obtained from integration of King
models, adopting the central surface brightnesses reported by H10.

\subsection{Integrated magnitudes from aperture photometry}
Direct measurements of the integrated magnitude were obtained by
performing AP using the {\it PHOT} task under the IRAF package.
For each cluster we used a
single fixed aperture with radius equal to the half-light radius
($r_{\rm h}$) quoted by MVM05 (or taken from H10 for the 5 clusters
mentioned above).  By definition, the total magnitude was then
computed by adding -0.75 to the value measured within $r_{\rm h}$.
The adopted center, background flux, and $E(B-V)$ values were the same
as above. Again the integrated instrumental magnitudes have been
converted to the ABMAG GALEX magnitude system and corrected for
extinction.  The photometric errors are defined as the standard
deviations of the fluxes. For both approaches, we ignored other
sources of errors like those that might affect reddening or distance
values. A $10\%$ error in the adopted $E(B-V)$ values may lead
to $\Delta {\rm mag}\sim0.3$ in both filters which in
turn gives $\Delta (FUV-NUV)\sim0.04$ or $\Delta (FUV[NUV]-V)\sim0.15$
for the most reddened clusters in our sample, such as NGC~6342 or
Palomar 11. 
Some of the clusters in our sample may be affected also by differential reddening
(see the case of NGC~6342; Alonso-Garcia et al. 2012). 
The impact of extinction variations was checked by means of synthetic experiments. 
We simulated a cluster of radius $R$ populated by 10,000 stars, all having the same observed 
magnitude in the three filters  ($FUV$, $NUV$ and $V$); we also assumed a mean $E(B-V)=0.2$. 
We then considered 10 circular areas of radius $0.1 R$ located at randomly extracted positions
within the cluster, and we increased by 0.2 the color excess of all the stars falling
within these areas; analogously, we considered 10 additional similar areas and we 
decreased by 0.2 the color excess of their stars: we therefore simulated a cluster with
mean $E(B-V)=0.2$, containing 10 bubbles with $E(B-V)=0.4$ and 10 regions with $E(B-V)=0$
(this corresponds to a differential reddening of amplitude $\Delta (E(B-V))=0.4$, 
as observed in NGC~6342 by Alonso-Garcia et al. 2012). We performed 1000 random extractions 
of the positions of these 20 areas, and, for each of them, we computed the integrated magnitudes
in every band and compared them to the input values. At the end of the experiments we 
found that the net color variation due to differential reddening is negligible. 
The largest effects ($0.02-0.03$ mag) is found  for the $(FUV-V)$ and $(NUV-V)$ colors.

\subsection{Comparison between SBP and AP integrated magnitudes and colors}
\begin{figure}[!hp]
\begin{center}
\includegraphics[scale=0.8]{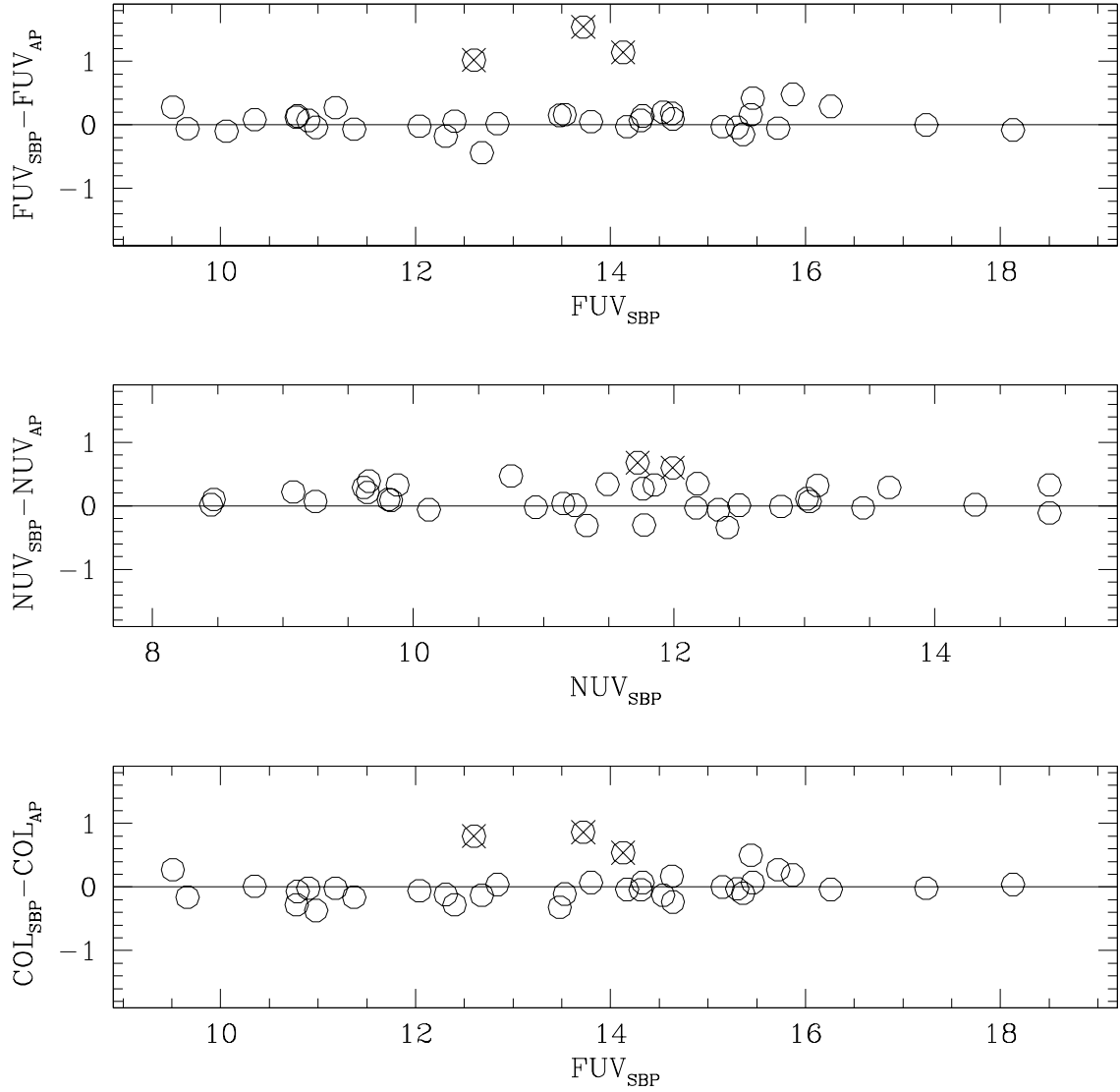}
\caption{Comparison between magnitudes and colors obtained by using
  surface brightness profiles fitting (SBP; Sect. 3.1) and aperture photometry (AP; Sect. 3.2).
   COL stands for
  (FUV-NUV).  The few clusters showing large differences are marked
  with large crosses. }
\label{figcomp}
\end{center}
\end{figure}

A comparison between the integrated magnitudes
obtained with the two methods is shown in Figure~\ref{figcomp}.  The
top panel shows that for most of the clusters there is a very good
agreement between the $FUV$ integrated magnitudes obtained from SBP
King fitting and from AP.  Most of the scatter ($rms=0.17$) is due to uncertainties
in the adopted half-light radii.  Three clusters (marked with black
crosses) are strongly deviant in the $FUV$ panel (at the top).  They are 47~Tuc,
NGC~1851 and NGC~6864 for which $(FUV_{SBP}-FUV_{AP})=1.02, 1.54$ and $1.14$ respectively.
Visual examination of the images reveals that
the large $\Delta FUV$ is due to the presence of very bright stars at
$r<r_{\rm h}$.  Caution must be paid in these cases, since these objects may be cluster
members as for the case of 47~Tuc (Dixon et al. 1995) or non-members as 
for NGC~1851 (Wallerstein et al. 2003). 
 In the middle panel of Figure~\ref{figcomp}, the $NUV$
integrated magnitudes are compared. The average difference is
essentially zero (with $rms=0.23$) and in this case the outliers are 
only NGC~1851 and NGC~6864 ($(NUV_{SBP}-NUV_{AP})=0.68, 0.60$). 
The bottom
panel shows the differences in the $(FUV-NUV)_0$ colors computed with
the two approaches as a function of $FUV_{\rm SBP}$.  Good agreement
is found also in this case ($rms=0.28$).

As discussed above, magnitudes based on SBP fitting 
are robust against stochastic effects due to the
presence of a few very bright stars. This is especially important in
the $FUV$, where UV-bright objects such as post-AGB and AGB-manqu\'e
stars are known to yield an important contribution to integrated light
(e.g., Paper I) thus significantly affecting the integrated colors of not resolved stellar
populations (see Figure~2).
However in the present work
we focus mainly on clusters global properties which are more 
likely traced by the bulk of hot stars.
 Therefore, in the following analysis we adopt SBP
colors, with the exception of the six clusters that could not be
  properly fitted by King models.

\section{The UV integrated colors}

\subsection{Dependence on HB morphology parameters}
\begin{figure}[!hp]
\begin{center}
\includegraphics[scale=0.8]{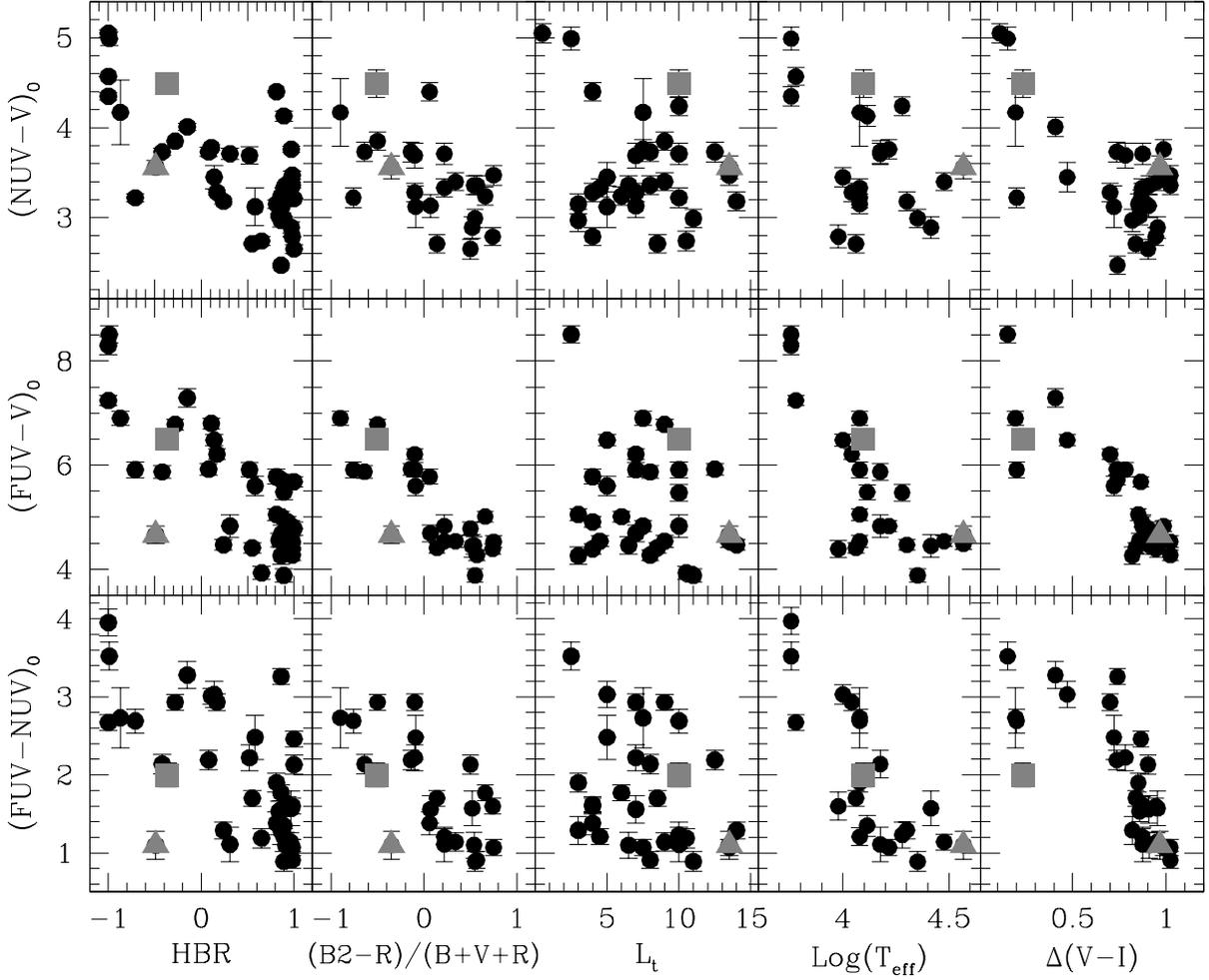}
\caption{UV integrated colors \emph{vs} five different horizontal branch 
  classification parameters. Starting from the left: HBR as proposed
  by Lee, Demarque \& Zinn (1994), $(B2-R)/(B+V+R)$ by Buonanno et al. (1993; 1997),
  horizontal branch length (L$_{\rm t}$) by Fusi
  Pecci et al. (1993), temperature extension Log($T_{max}$) as
  introduced by Rec\'\i o Blanco et al. (2006), and median horizontal branch color
  $\Delta(V-I)$, by Dotter et al. (2010). The grey triangle and square 
  are NGC~2808 and NGC~1851 respectively; these have bimodal horizontal branches}
\label{multi}
\end{center}
\end{figure}

A number of HB morphology classifications have been proposed over the
decades.  We analyze here the behavior of UV colors as a function of
the most important HB parameters.  Perhaps the most commonly used is
$HBR =(B-R)/(B+V+R)$ introduced by Lee, Demarque, and Zinn (1994),
where $V$ is the number of variables, and $B$ and $R$ are the numbers
of HB stars blue-ward and red-ward of the instability strip. In the
leftmost panels of Figure~\ref{multi} the UV colors are shown as a
function of $HBR$. The large number of clusters with $HBR \sim 1$
simply reflects our target selection as most clusters with red HBs are located 
at low Galactic latitude. The wide range of colors at any
given value of $HBR$ arises because there is a large variety of HB
morphologies even among the subset of clusters with predominantly blue
HBs. The $HBR$ parameter is insensitive to the details of the color distribution of stars bluer than the RR
Lyrae. The same applies to other HB morphology parameters defined on the basis of optical CMDs.
Some clusters with bimodal HBs, like NGC~2808 (plotted as a gray
triangle) and NGC~1851 (plotted as a gray square), have an $HBR$ value
that ranks them among clusters with reddish HBs, yet they are ``hot''
in the UV.

We have also compared our UV colors with the HB parameter $(B2-R)/(B+V+R)$ defined by 
Buonanno et al. (1993; 1997),
where B2 is the number of stars bluer than $(B-V)_0 = -0.02$ (see also Catelan 2009).
Data have been taken from Buonanno et al. (1997) and Preston et al. (1991).\footnote{Some caution 
should be paid when using these ratios, because some differences
may come out if using more recent and deeper photometry especially for clusters with extended blue
tails.}
This parameter 
is expected to correlate more clearly with UV colors, since it is able to remove the degeneracy which 
characterizes HBR for clusters with extended blue HBs. 
In fact we find that both in $(FUV-NUV)_0$ and $(FUV-V)_0 $
there is a clear trend for clusters with $(B2-R)/(B+V+R)>-0.25$, in agreement with findings by Catelan 2009 
and theoretical expectations (Landsman et al. 2001).
Clusters with $(B2-R)/(B+V+R)<-0.25$ are those in the 
"Bimodal HB" zone (Catelan 2009). In this region in fact, we find NGC 1851 and NGC 2808, as well 
as other clusters with
a clear bimodal HB like NGC 6864 (M75) and NGC 7006, or with a more populous red HB and a 
sparsely populated blue HB as NGC~1261 and NGC~362. In the latter case the extension
to the blue of the HB 
may be actually due to contamination by background 
stars belonging to the Small Magellanic Clouds. 
A Spearman correlation rank test gives probabilities larger than $99.95\%$ for
correlations with both $(FUV-NUV)_0$ and $(FUV-V)_0$. 
In the $(NUV-V)_0$ the correlation is less clear, in particular if bimodal clusters 
are not considered. 
In this case the Spearman test gives a probability P$\sim85\%$.  
The $(B2-R)/(B+V+R)$ parameter is more efficient than $HBR$ in characterizing 
 extended blue HBs. Thus, in general, it would be preferable. However we stress that $(B2-R)/(B+V+R)$ 
 may suffer of the same limitations as $HBR$ since it is defined and measured in optical CMDs
 where star counts along the extreme HB blue tails could be significantly incomplete. 
 We therefore encourage the definition of similar quantities based on a proper combination of UV
 and optical bands.

\citet{fpbt} defined several parameters describing HB blue tails (BTs)
using photometry in the optical bands. In principle, their parameter
L$_{\rm t}$ describing the length of BTs should correlate well with
the UV light output of GCs. However, as shown in the third column of
panels in Figure~\ref{multi}, there is no obvious correlation. The data
used by \citet{fpbt} were not uniform and in a few cases they were based on CMDs
dating back to the 1960's. In many cases the CMDs were not deep enough
to show BTs that we now know to exist. This has also been noticed by
\citet{gratton10}.

\citet{recio06} included BTs in their analysis of HB morphology.
They used a homogeneous deep HST survey made in the $F439W$ and
$F555W$ (roughly $B$ and $V$) filters. They measured the length of BTs in
terms of the maximum effective temperature \recio\ reached by the HB: the values of
\recio\  were obtained by comparing in the CMDs the observed BTs with
theoretical HB models. However, the optical plane is not ideal
for determination of the temperatures of the hottest stars.
For this reason, this parameter should be considered as a lower limit
of the real HB extension, especially for the most extended BTs where
the HB may be truncated in optical CMDs. 
Indeed, incompleteness may strongly affect clusters with a population of very hot 
HB stars like the Blue Hook stars. These objects have been suggested to 
be stars which experienced the helium-flash at high effective temperatures (see for example Moehler et al. 2004;
Busso et al. 2007; Rood et al. 2008; Cassisi et al. 2009; Brown et al. 2010; Dalessandro et al. 2011; 
Brown et al.2012)
or stars with an extremely large helium mass fraction ($Y>0.5$; D'Antona et al. 2010).
However even in those cases in which incompleteness does not represent a limit,
optical photometry still remains a poor measure of \recio\ for extremely hot stars.
In fact, as shown for example in
Dalessandro et al. (2011) in the case of NGC~2808, the value of
$T_{eff}$  by \citet{recio06} is underestimated by $\sim 10,000$\,K.
However even with this limitation, \recio\ is a useful parameter
describing the UV bright population of GCs \citep[see,
  e.g.,][]{gratton10}.
The fourth column of panels in Figure~\ref{multi} show the UV colors as
a function of \recio\ . There is an obvious correlation in the case of
colors involving $FUV$ magnitudes, not so much in the case of
$(NUV-V)_0$. This is expected, since NUV magnitudes are less sensitive than FUV
to the \recio\ of the hottest stars. A Spearman test gives probabilities larger than $99.99\%$
for correlations with both $(FUV-NUV)_0$ and $(FUV-V)_0$, and $\sim92\%$
for the correlation with $(NUV-V)_0$.  All the clusters with $\recio\ \ga
4.2$ (roughly 1/3 of the sample) have approximately the same
$(FUV-NUV)_0$ color. There is another group of clusters with $4.0 \la
\recio\ \la 4.2$ with a wide range of $(FUV-NUV)_0$.  Therefore, although
the correlation between this parameter and integrated color is
statistically robust for colors involving $FUV$ magnitudes, the
detailed dependence of integrated color on HB morphology as defined by
this parameters seems not to be monotonic.

\citet{dotter10} introduced the parameter \dotter\ to describe HB
morphology. It is defined as the difference in the median colors of the
HB and the red giant branch (RGB) at the level of the HB, and it is
derived from a homogeneous, deep HST ACS survey in F606W and F814W
(roughly $V,~ I$). However, since BTs are almost vertical 
in the $(V,V-I)$ CMDs, we might expect that \dotter\ has
problems similar to $HBR$ in characterizing the GC UV light.  The
rightmost column in Figure~\ref{multi} shows the UV colors as a
function of \dotter. The correlations are good, in a statistical sense:
probabilities larger than $99.99\%$ are found for $(FUV-NUV)_0$ and
$(FUV-V)_0$ colors, while $\sim 93\%$ probability is obtained for a
correlation with $(NUV-V)_0$. However the bulk of the sample is bunched
up to the lower right corner of the plots.  This is because,
unsurprisingly, a parameter based on red optical colors
distinguishes clusters with relatively red HBs from their blue counterparts, 
but provides very poor
discrimination between clusters with blue and extremely blue HB morphologies.

\subsection{Dependence on metallicity}
\begin{figure}[!hp]
\begin{center}
\includegraphics[scale=0.8]{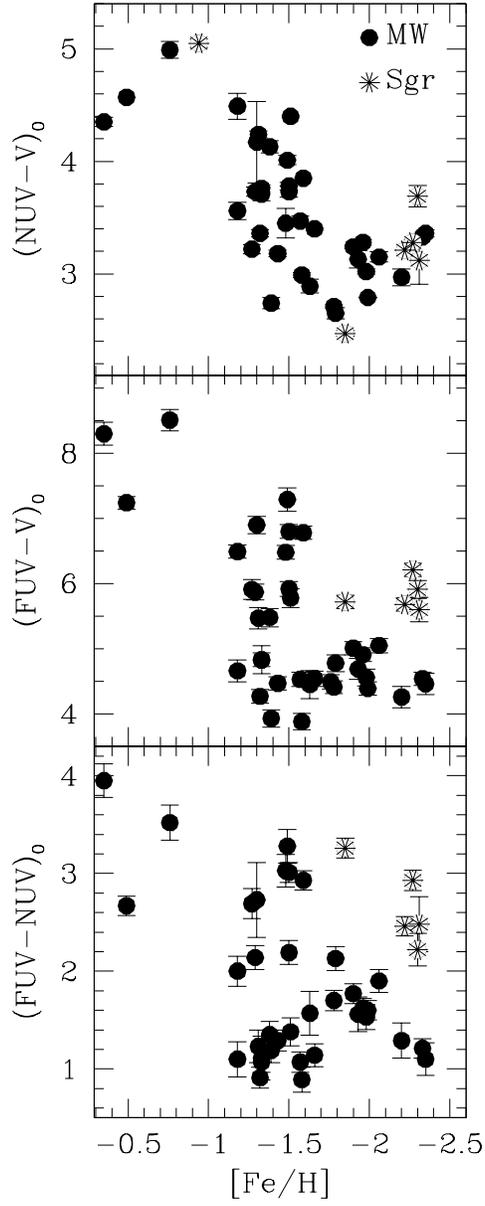}
\caption{UV integrated colors as a function of metallicity in Carretta et al. (2009b)
  scale. Clusters possibly connected with the Sagittarius stream are
  plotted as asterisks. }
\label{met}
\end{center}
\end{figure}

In order to investigate any possible link between the UV colors and
chemical compositions of the Milky Way GCs, we adopted the
\feh\ values quoted by Carretta et al. (2009b). For
Terzan~8, which is not in their sample, we used 
the equation listed by Carretta et al. (2009b) 
to convert \feh\ values from Zinn \& West (1984) to their metallicity scale.
The UV colors derived from SBP fitting are plotted as a function of
metallicity in Figure~\ref{met}.

First focusing on the top panel of Figure~\ref{met}, one can see that,
for the GCs included in this sample, $(NUV-V)_0$ decreases by about 2
magnitudes as \feh\ decreases from $\sim$ -0.7 to $\sim$ --1.5. For
smaller values of \feh, $(NUV-V)_0$ is roughly constant, or perhaps
increases slightly as \feh\ decreases.  This clear, although
non-monotonic, trend of $(NUV-V)_0$ with metallicity is confirmed by a
Spearman correlation rank test, according to which the probability of
a correlation is 99.99\% ($>4\sigma$).  Even after removing the most
metal-rich clusters (${\rm \feh}>-1$) the probability of correlation
remains significant at more than $4\sigma$.  By removing from the
sample the most metal-poor clusters ($\feh<-1.5$) the probability for
a correlation actually increases.  The more restrictive non-parametric
Kendall rank correlation test also shows that there is
a strong and positive correlation between $(NUV-V)_0$ and \feh\, with
a significance of $\sim4.1\sigma$.
The integrated $NUV$ radiation from GGCs is dominated by blue HB stars
but it has some contribution also from turnoff and blue stragglers
stars (see for example Figure~1 in Ferraro et al. 2003).
 The overall trend of $(NUV-V)_0$ as a function of metallicity
seen on the top panel of Figure~\ref{met} is therefore in line with
expectations from standard stellar evolution theory, as HB and turnoff
stars in more metal-rich clusters are expected to be cooler and
redder.  This expectation was also confirmed by observations collected
with ANS, OAO, and UIT, as presented by DOR95.  Metallicity, however,
is only one of the parameters determining the color distribution of HB
stars.  At least one {\it second parameter} is known to affect the
color of the HB in GCs, and it can be recognized in the large spread in
color for $-1.5\simless$ [Fe/H] $\simless -1.0$, where clusters with
similar \feh\ can have $(NUV-V)_0$ differing by as much as $\sim$ 1.5
mag.  The nature of this second parameter has been the subject of
debate for several decades now, with candidates such as He abundance,
age, mass loss, binarity, rotation, among others, being suggested in
the past to explain the effect (for a review, see Catelan 2009).
Recently Dotter et al. (2010)
and Gratton et al. (2010) have particularly emphasized the role of age 
and they also stressed on the necessity of
even a third parameter  (the central luminosity density for the first and
the He abundance for the second). 

Interpretation of the dependence of colors involving $FUV$ magnitudes
as a function of \feh\ requires a little more care. At first glance,
the middle and bottom panels of Figure~\ref{met} show a less
clear correlation between color and \feh.  The bottom panel in
particular looks like a scatter plot.  It is true, though, that,
$(FUV-V)_0$ varies by almost 5 magnitudes.
Indeed, a Spearman test gives a probability
$P\sim99\%$ (corresponding to $\sim 2.3 -2.5\sigma$) that $(FUV-V)_0$
is correlated with metallicity.  The Kendall test gives correlation
probability of $\sim2.1\sigma$.  The probability drops to $\sim80\%$
if the most metal-rich clusters are excluded from the analysis.  At
face value, therefore, $(FUV-V)_0$ correlates with \feh\ in a
similar way as $(NUV-V)_0$, although in a less strong or noisier
fashion.

In summary, then, one finds that the behavior of GGCs in the
$(FUV-V)_0$--\feh\ and $(NUV-V)_0$--\feh\ planes is essentially the
same.  On both planes, three sub-families of clusters can be
recognized: 1) GGCs with $\feh\simgreater -1.0$, which are predominantly red
\footnote{This is partially due to the incompleteness of our sample, as it does not include
NGC 6388 and NGC 6441, which are metal-rich but have an EHB extension};
2) GGCs with $-1.5 \simless$ $\feh\simless -1.0$, the
``second parameter region'' (see also Fusi Pecci et al. 1993),
 where GGCs have a wide range of colors,
about $\sim 2$ mag in $(NUV-V)_0$ and $\sim 4$ mag in $(FUV-V)_0$; and
3) GGCs with $\feh\simless -1.5$, which are all blue.  It is worth
noticing that, intermediate-metallicity ([Fe/H]$\approx-1.5$) clusters
are the bluest in the three colors combinations. This was also
highlighted by DOR95. The extension of their HBs (see Paper I) is
compatible with their integrated colors. On average, the metal-poor
($\feh\simless-1.7$) GCs have redder HBs than the intermediate ones.
This appears to contradict expectations based on the notion that metallicity is  
the {\it first parameter}. 
Some authors interpreted this discrepancy invoking age differences among clusters.
However Dotter (2008) was able to account for this behavior without invoking 
age differences, by using
synthetic HB models and simple assumptions about the
mass-loss/metallicity relation.

\subsection{GCs in the Sagittarius stream}
\label{sec:sagit}
Careful inspection of the $(FUV-NUV)_0$ or $(FUV-V)_0$ \emph{vs}
\feh\ plots in Figure~\ref{met} clearly reveals that the color spread
at $\feh<-1.5$ is due to a subset of clusters (plotted as asterisks),
which are systematically redder by $\sim$ 1.5 and 1.0 mag
in $(FUV-NUV)_0$ and $(FUV-V)_0$, respectively, than the other GCs in
the same metallicity regime. This is also confirmed if colors obtained
with AP are used instead of those derived from SBP fitting. Likewise,
adopting the Zinn \& West (1984) metallicity scale or metallicity values 
by Carretta \& Gratton (1997) does not affect the general behavior, even
though minor differences in a cluster-to-cluster comparison can
obviously come out.  Interestingly these clusters (NGC~4590, NGC~5053,
NGC~5466, Arp~2 and Terzan~8) are potentially connected with the
Sagittarius dwarf galaxy stream (Dinescu et al. 1999, Palma et
al. 2002, Bellazzini et al. 2003, Law \& Majewski 2010), and thus may
have an extra-Galactic origin.  We stress that all these clusters have
been suggested to be connected to the Sagittarius stream by at least
two different authors. The other candidate Sagittarius GC is the
relatively metal-rich ([Fe/H]=$-0.94$) Palomar~12, for which we were not able
to get $FUV$ magnitude (see Section~3.1). Among the clusters considered here, 
the classification of NGC 4590 is the most uncertain. According to its metallicity, HB morphology and
proper motions (Smith et al. 1998; Dinescu et al. 1999; Palma et al. 2002) 
it is likely associated with the Sgr stream, although Forbes \& Bridges (2010)
suggest that it is connected to the Canis Major dwarf. 
In any case, the extragalactic origin seems to be well established.

\begin{figure}[!hp]
\begin{center}
\includegraphics[scale=0.8]{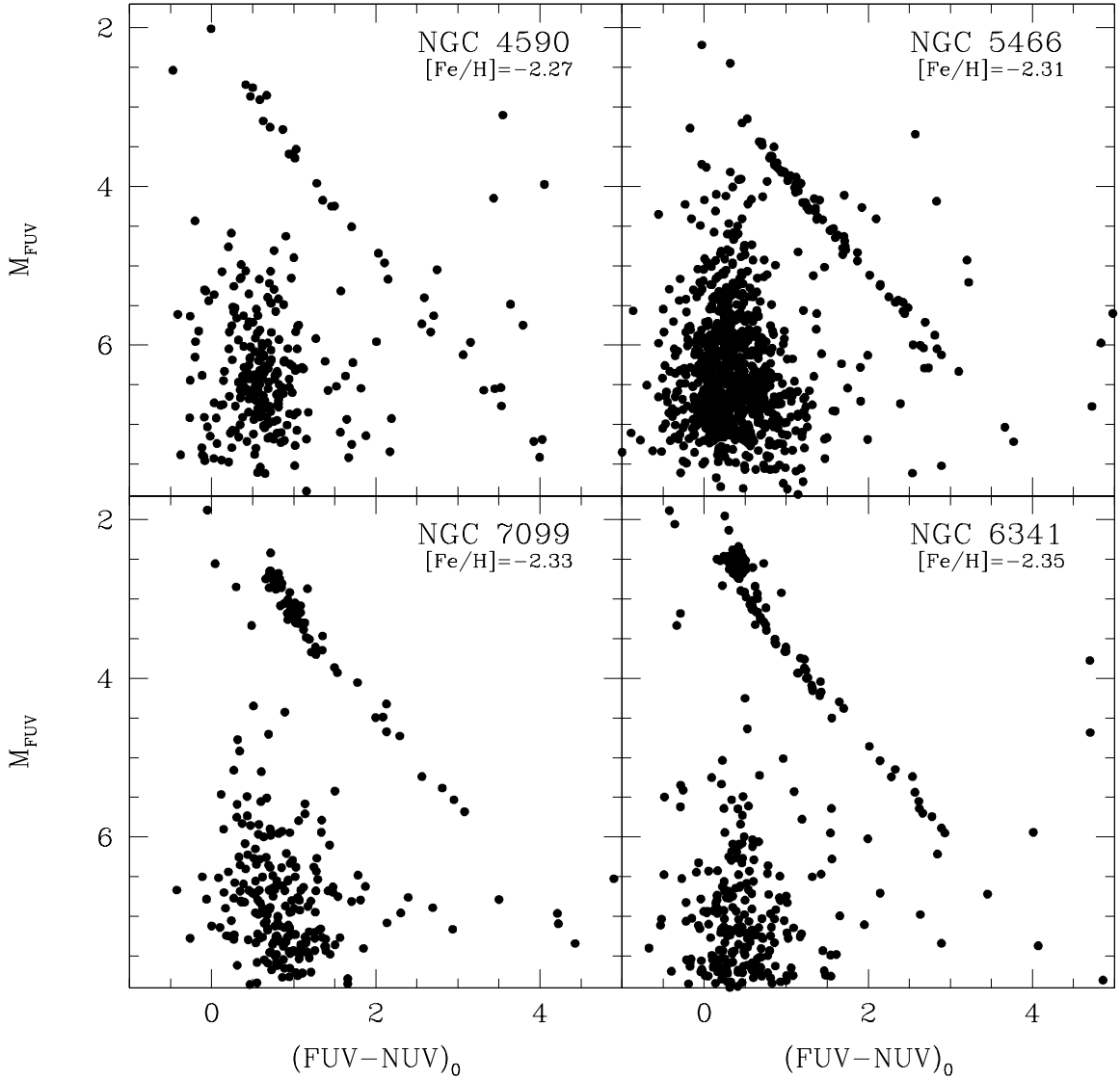}
\caption{Comparison between GALEX CMDs of two GCs connected with
  the Sagittarius stream (upper panels), and GGCs in the same
  metallicity regime (lower panels).}
\label{sgr_cmd}
\end{center}
\end{figure}

To explore the significance and meaning of this behavior we used the
photometric catalogs presented in Paper I to understand how the differences 
in the integrated colors relate to
differences in the cluster CMDs. We picked two clusters associated
with the Sagittarius stream, and compared them with GGCs of similar
metallicity (Figure~\ref{sgr_cmd}). The photometric catalogs have been
corrected by distance modulus and reddening values reported by H10, as
done in Paper I.  The HBs of the stream clusters are homogeneously
populated in color and
magnitude. In contrast, the GGCs show an obvious increase of star
density towards higher temperatures and bluer colors.  In these
clusters the median color of HB stars is bluer by
$(FUV-NUV)_0\sim1$.  This difference is qualitatively consistent
with the measured discrepancy between the integrated colors. The
HB star distributions differences in the GALEX CMDs are 
in agreement with results from
optical high-resolution surveys (see Dotter et al. 2010 for example).
The Sagittarius GCs do not show any systematic trend in the
$(NUV-V)_0$ {\it vs} [Fe/H] diagram. More details about the morphology
of their HBs will be presented in Paper III (R. T.  Rood.  et al.,
2012 in preparation). 

We checked also for possible differences in relative ages.  For
this, we used two independent papers (Salaris \& Weiss 2002, and
Dotter et al. 2010) containing age estimates for a large sample of GCs
including most of our targets.  Even though some systematic
differences are present between the two age scales, within the
uncertainties the clusters connected with the Sagittarius stream are
classified in both cases as old and coeval with genuine GGCs in the
same metallicity regime, with the only exception of Pal~12 which is
$\sim3-5$ Gyr younger.  In particular, the four Sagittarius old
clusters in common with Salaris \& Weiss (2002; Terzan~8 has not been
analyzed by the authors) have an average age of $(11.4 \pm 0.6)$
Gyr which is fully compatible with the estimates for GGCs.
In Dotter et al. (2010) the differences between the Galactic and
the Sagittarius GCs are of the order of $\sim 0.5$ Gyr.

In addition, recent high resolution spectroscopic analysis (Carretta
et al. 2010) showed that, on average, the Sagittarius clusters in our
sample share the same $\alpha$-elements abundances with their Galactic
twins. We used the {\it R'-parameter}
reported by Gratton et al. (2010) to highlight possible differences.
The {\it R'-parameter} is defined as the ratio
between the number of HB stars and that of RGBs brighter than the level corresponding
to $V_{\rm HB}+1$.
This quantity is an indirect estimate of $Y$ (see Cassisi et al. 2003; Salaris et al. 2004), 
since the HB luminosity, as well as the RGB and HB lifetimes, depends on the He content.
However for this comparison we preferred to use only the {\it R'-parameter} 
in order to avoid uncertainties that may come from the calibrations used to 
derive helium abundances ($(Y(R')$) from it.

Three (NGC~4590, NGC~5053 and NGC~5466) out of the five
Sagittarius clusters have been studied by Gratton et al. (2010). 
It is interesting to note that these clusters have {\it R'} values
smaller than other clusters with similar metallicity [Fe/H]$<-1.5$.
Given the statistical uncertainties of these measurements, the difference in {\it R'} between
any two given clusters is somewhat uncertain.
Therefore we performed a t-test to check the significance of the difference between
the mean values of the
two distributions. We find that for the clusters potentially connected with the Sgr 
stream $<R'>=0.48\pm0.01$ while for GGCs $<R'>=0.74\pm0.18$. 
The t-test gives a probability P$>99.9\%$ that they are different.
We stress that it is not possible to make a final statement about this point. 
In fact this simple analysis suffers
of low number statistics and incompleteness of our sample. Moreover the significance of the result 
may depend on the clusters considered to be connected with Sagittarius.
However it emerges that clusters connected with Sagittarius share, on average, the same
properties as the genuine GGCs, except for the {\it R'-parameter}.
This difference might be an indication that those 
clusters have lower He abundances than GGCs in the same metallicity regime, and this is likely the 
main responsible of the differences in FUV integrated colors.
Further analysis and estimates of the {\it R'-parameter} for the other clusters possibly associated with
Sagittarius are highly desirable and could provide stronger constraints on this problem.
 
\subsection{Dependence on other GC properties}
\begin{figure}[!hp]
\begin{center}
\includegraphics[scale=0.8]{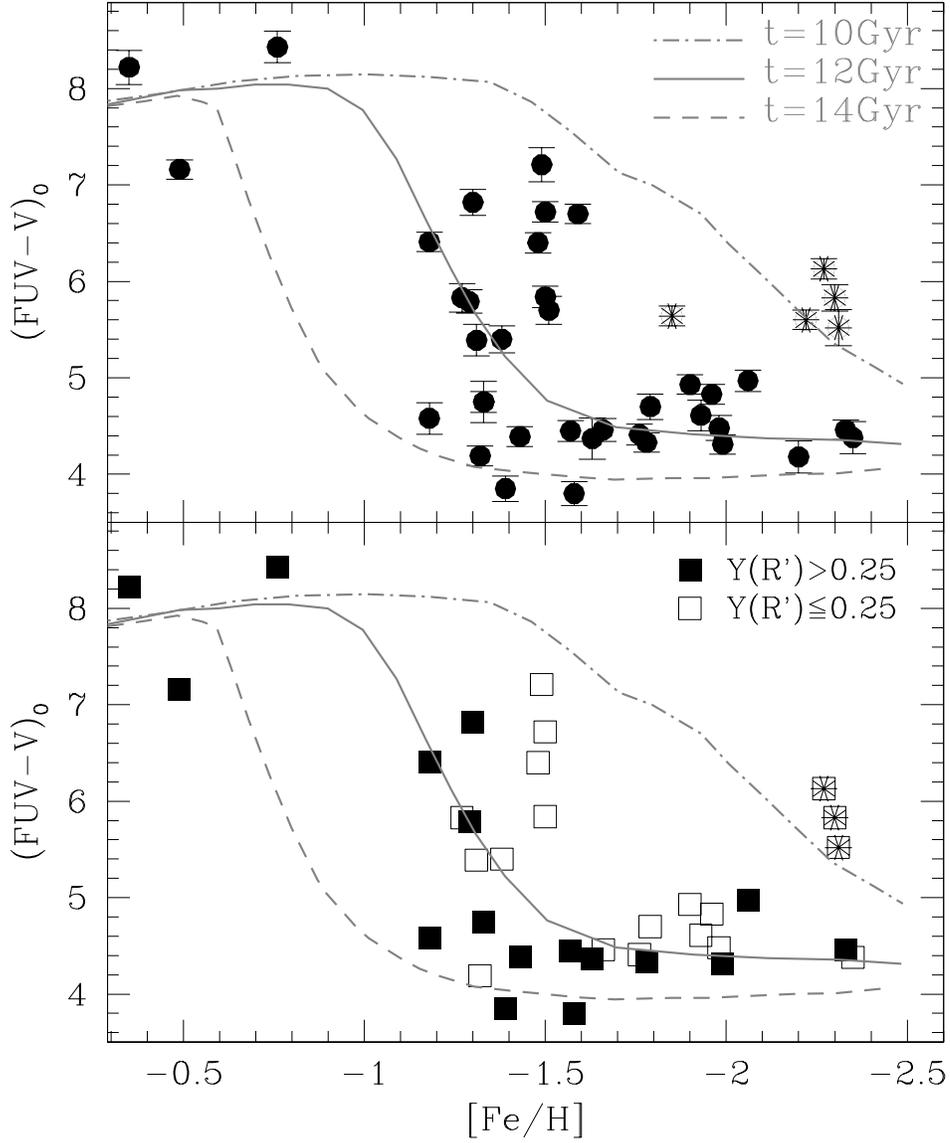}
\caption{{\it Top Panel}. $(FUV-V)_0$ color as a function of metallicity as in
  Figure~\ref{met}, with superimposed theoretical models by Lee et
  al. (2002). {\it Bottom Panel}. The cluster sample has been split according to
  the helium content: black squares are ``Helium-rich'' clusters
  ($Y(R')>0.25$), while the open squares are the ``He-poor''
  ones. Only clusters in common with Gratton et al. (2010) are plotted. 
  As before, asterisks are clusters potentially associated with the Sagittarius stream.}
\label{met_mod}
\end{center}
\end{figure}

A number of authors (Lee et al. 2002, Sohn et al. 2006, Rey et
al. 2007) made use of UV integrated colors to investigate age trends
in the GC systems of the Milky Way (by using OAO~2, ANS and UIT data), as well
as of M31 and M87.  Lee et al. (2002) and Yi et al. (2003) argued that
with a proper modeling of the HB contribution to UV bands, the
$(FUV-V)$ could be a good age tracer of relatively old stellar
populations. The basic assumption is that the HB morphology is driven
by metallicity and age.  Of course a number of uncertainties may arise
because of the treatment of mass loss along the upper RGB, or the
presence of hot HB stars, like EHB, which are not included in these
models. However under this assumption, both Lee et al. (2002) and Rey et
al. (2007) were able to reproduce the UV color \emph{vs} metallicity
distribution of Milky Way and M31 GCs. In contrast, Sohn et al. (2006) were
unable to find an acceptable match between their data for the M87 GCs
and the models of Lee et al. (2002), which required unphysically old
ages ($t\sim16\,\gyr$) to reproduce the observed color distributions.
Taking advantage of our large and homogeneous sample and the high
photometric accuracy of our data, in Figure~\ref{met_mod} we compare
the $(FUV-V)_0$-[Fe/H] distribution observed in our GC sample,
with the theoretical models of Lee et al. (2002) for integrated colors
of SSPs with different ages.  The color-metallicity distribution of
GGCs is consistent with ages $10 <t< 14$ Gyr, in broad agreement with the
results obtained from fitting theoretical isochrones to the turnoff or
the white dwarf cooling sequences (e.g., Salaris \& Weiss 2002; Dotter
et al. 2010).  It is worth noticing that the three clusters that 
appear as the "youngest" ($t\sim 10$ Gyr), according to the Lee et al. 2002 models, 
are those connected with the Sagittarius stream (NGC~4590, NGC~5053 and NGC~5466).
As discussed above,
however, they most likely are old and coeval with those in the same
metallicity range, thus indicating that caution must be used to derive
ages from this color-metallicity plane (Fusi Pecci et al. 1993), since other 
parameters play a role in distributing
clusters in this diagram. The helium content certainly is one of
these parameters (see for example Sohn et al. 2006, Kaviraj et al. 2007,
 Chung et al. 2011).

Indeed, there is now evidence that at least a few massive GCs could
host multiple populations with different He abundance (see Piotto 2009
for a review).  As recently shown by Chung et al. (2011; see also
Kaviraj et al. 2007), helium might have a strong impact on the UV
emission from an old stellar population. They suggest, in fact, that
He-rich sub-populations in GCs could be able to reproduce the
``UV-upturn'' observed in elliptical galaxies. To check the impact of
multiple populations on the observed colors we used the helium
fractions $Y(R')$ derived by the {\it R'-parameter} by Gratton et al. (2010) for the 36
clusters in common.
We split our sample using $Y(R')=0.25$ as
threshold, to obtain two roughly equally populated
sub-samples, with mean $Y(R')$ of about 0.27 and 0.23.\footnote{We stress that this is only a broad 
selection aimed at obtaining two equally populated sub-samples. 
As already pointed out in Section 4.3, $Y(R')$ may suffer from calibration uncertainties.
In this case however, we were forced to use it since {\it R'} shows a mild trend with 
metallicity (Gratton et al. 2010), which would have prevented us to 
make a comparison for the entire sample.}
From Figure~\ref{met_mod}, it is evident that such differences in He content can have effects on UV
colors equivalent to age differences of $\sim2Gyr$.

\begin{figure}[!hp]
\begin{center}
\plotone{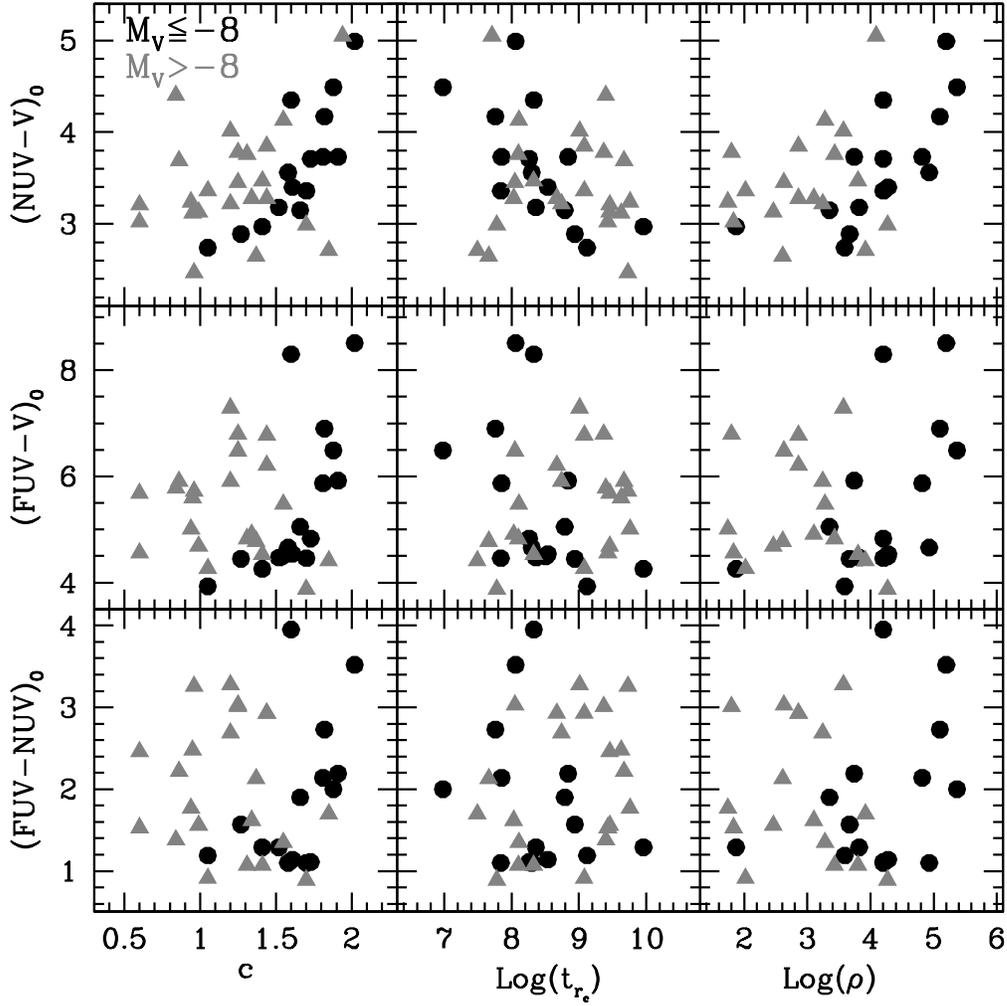}
\caption{GALEX UV colors plotted as a function of cluster
  concentration, central relaxation time at $r_c$ and central density. 
  The sample has been split in two sub-samples: more
  massive clusters ($M_V\leq-8$) are shown with black circles, less
  massive clusters are plotted as grey triangles. The four post-core
  collapse clusters in our sample are not plotted since the values of
  their concentration ($c=2.5$) is just arbitrary.}
\label{concentration}
\end{center}
\end{figure}

While a detailed discussion about the long standing ''HB second
parameter'' problem is postponed to Paper III, here we briefly
consider how UV integrated colors vary with clusters parameters
previously suggested to be connected with the presence of long HB BTs.
\citet{fpbt} suggested that BTs are related to central density and
concentration, and that stellar interactions somehow enhanced mass
loss. \citet{dotter10} also suggested that the third parameter acting
in modeling the HB morphologies is somehow related to the chance of
interactions between stars.  Figure~\ref{concentration} shows the
GALEX colors plotted as a function of concentration $c =
\log(r_t/r_c)$, where $r_c$ is the King model core radius and $r_t$ is
the tidal radius. While the sample as a whole shows no significant
correlation, the massive cluster ($M_V >-8$) colors are well
correlated with $c$, moving to the redder (or cooler) colors as the
concentration increases. The Spearman test gives $97.5\%$ probability
of correlation between $c$ and $(FUV-NUV)_0$, larger than $99.95\%$ with
$(FUV-V)_0$ and about $99.8\%$ with $(NUV-V)_0$. This is the opposite of
what we had expected from Fusi Pecci et al. (1993) and Dotter et al. (2010)
interpretations. Similar results are obtained, although
with lower significance, for the relaxation
time and central density (Figure~\ref{concentration}).
However a direct comparison with results by Dotter et al. (2010) could not be performed, since in their
analysis the authors removed the effects of metallicity and age from their parameter \dotter\ .
We defer to Paper III for a detailed and comparative analysis.

\section{Comparison with GCs in M31 and M87}
\begin{figure}[!hp]
\begin{center}
\includegraphics[scale=0.8]{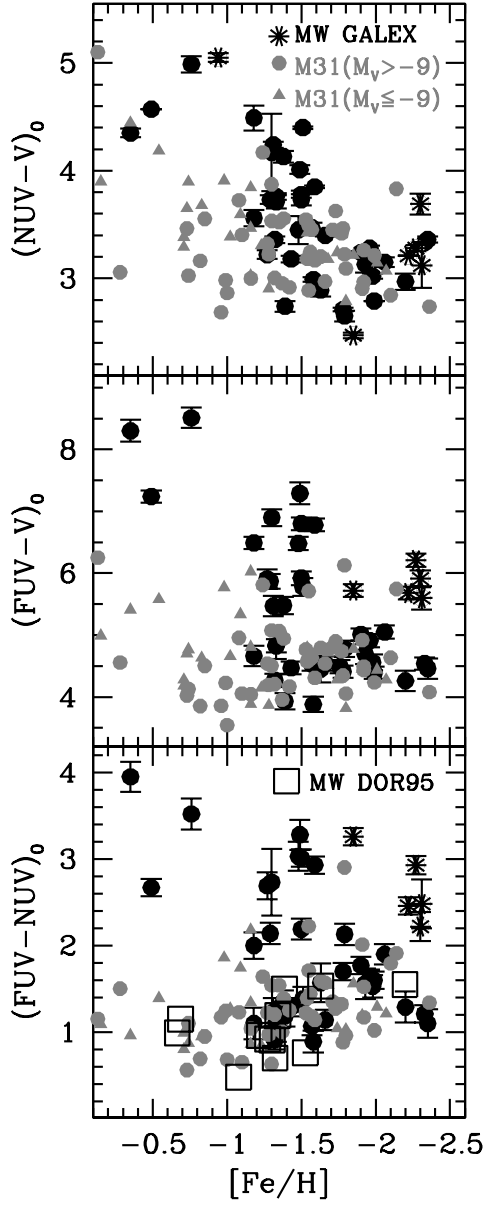}
\caption{GALEX colors of our GGC sample (black) compared to the GCs in
  M31 (from Kang et al. 2011). The most massive M31 clusters (with
  $M_V\leq-9$) are plotted as grey triangles, while the less massive
  ones as grey circles.  In the lower panel our GGC sample has been
  supplemented with clusters observed with ANS and OAO2 by DOR95 (open
  squares).}
\label{m31comp}
\end{center}
\end{figure}

\citet{galexm31} have used GALEX to observe GCs in M31. 
The authors (Kang et al. 2011) have recently published an updated catalog with a larger
sample of clusters and reviewed reddening values.
In Figure~\ref{m31comp} we compare our results with data for M31 clusters 
classified as "old" ($t>2 Gyr$) by Caldwell et al. (2011) and 
we restrict the sample to clusters with 
$E(B-V)<0.16$, in order to avoid clusters with high reddening uncertainties.\\
M31 seems to show a lack of red clusters with respect to the Galaxy.
However, this is likely due to the limited sensitivity of GALEX to detect
relatively red populations in distant systems (see discussion in Rey et. al. 2007).
Hence for the comparison we focus on the bluest systems,
$(FUV-NUV)_0\lsim1.5$, $(FUV-V)_0\lsim5$ and $(NUV-V)_0\lsim3.5$,
which, for the GGCs, correspond to a metallicity range $-2.5 < \feh
< -1.0$.  In this metallicity regime the distributions in the Milky Way and
in M31 are quite similar.  The bluest colors reached
are essentially the same, and the distributions show little variations
with metallicity. 
Three M31 clusters lie almost in the same 
region as GCs connected with the Sagittarius dwarf in the $(FUV-V)_0$-[Fe/H] plane.
They are G~327, B~366 and interestingly B~009 that is known to be projected 
against the dwarf spheroidal galaxy NGC~205. Unfortunately CMDs are available only for 
B~366 (Perina et al. 2009). This is an old cluster with a red HB and possibly
a very mild blue extension, which makes it compatible with its redder UV colors.

The case is very different at higher metallicity, $\feh>-1$. In the Milky Way
sample there are only red GGCs, while in M31 there are many blue
GCs.\\ 
In order to
make the comparison with M31 GCs as complete as possible, we have
supplemented our GALEX sample with 12 additional GGCs\footnote{NGC~5139, NGC~6093,
NGC~6205, NGC~6266, NGC~6388, NGC~6441, NGC~6541, NGC~6626, NGC~6681, NGC~6715, NGC~6752, NGC~7078} 
 from DOR95,  not observed by GALEX because
of the target selection limitations discussed in Section 2.  
First, we converted the ANS and OAO~2 magnitudes from the STMAG to the ABMAG system
and we checked that no
systematic color offsets or trends are present for the (15) GCs in
common. The average difference results to be $\Delta(FUV-NUV)_0=0.08$ and is
fully consistent with the reported errors. Then, we adopted the ANS
and OAO~2 magnitudes given in Table~1 of DOR95 and we corrected these
values by using the color excess $E(B-V)$ quoted by H10.  So as not to
introduce additional errors through the $V$ magnitudes, we consider
only the $(FUV-NUV)$ color.  The results are shown as open squares in
the lower panel of Figure~\ref{m31comp} and demonstrate that (at
least) two GCs, namely NGC~6388 and NGC~6441, in the
Galaxy have colors comparable to those of M31 at $\feh>-1$. Still, M31
appears to have many more hot metal-rich GCs. Why would that be the case?
As shown in
Figure~\ref{m31comp} roughly half of the blue, metal-rich M31 GCs are
indeed quite massive ($M_V \leq -9$).  Hence, the relative paucity of
hot, metal-rich GCs in the Milky Way could be due in part (but only in part) to
the fact that there are only two massive metal-rich clusters in our
supplemented sample. It is also possible that many GGCs with high metallicity 
and a blue HB are missed because of their location towards highly extinguished 
regions of the Galaxy.\\

We also compare GGC colors measured with GALEX with those obtained for
the giant elliptical galaxy M87 
using HST STIS images (Sohn et al. 2006). We converted those magnitudes
from the STMAG to the ABMAG photometric system.  In order
to perform a direct comparison, we transformed \feh\ values to the
metallicity indicator $Mg_2$ using equation A1 in the Appendix of Sohn
et al.  (2006).  As shown in Figure~\ref{m87}, M87 GCs are on average
bluer by $\sim1.5$ mag both in $(FUV-NUV)_0$ and $(FUV-V)_0$, while
they do not show any appreciable difference in $(NUV-V)_0$\footnote{Out of the total
sample of 162 clusters, 153 have $FUV$ magnitudes, 16 $NUV$  and 
only 7 have both.}. These
differences are consistent with what observed by Sohn et al. (2006) in
comparison with the DOR95 sample.  On the basis of what we discussed in
Section \ref{sec:sagit}, we may suppose that M87 GCs are on average
older or have a higher He content than the Milky Way objects.  As noted
above, the age-metallicity grid of theoretical predictions provides
realistic ages for the M87 GCs only  when the effect of Helium is taken into account
(see Figure 2 in Chung et al. 2011; Kaviraj et al. 2007).

From the comparison between GGCs and those belonging to three other
galaxies (the Sagittarius dwarf, M31 and M87), different behaviors
emerged. In fact the clusters associated with the Sagittarius dwarf are
 on average redder than
the MW ones, the M31 clusters have colors which are comparable to
those of the GGCs, while the M87 star systems are bluer. 
We note that there may be a possible trend between the mass 
of the host galaxy and
the color distribution of its globulars, in the sense that the higher is
the galaxy mass, the bluer are the GC UV colors.  In fact M87 (with
the bluer systems) is a super-giant elliptical that is about two orders of 
magnitude more massive than the Milky Way
\citep[$1.7\times10^{13}<M/M_{\odot}<4.0\times10^{13}$;][] 
{m87mass},
 while Sagittarius \citep[$\sim1.6\times10^8 M_{\odot}$;][]{sgrmass} with the reddest sample of GCs
(although quite small), is a dwarf galaxy, and M31 
\citep[$3.7\times10^{11}<M/M_{\odot}<2.5\times10^{12}$;][]{m31mass} 
and the Galaxy \citep[$2.4\times10^{11}<M/M_{\odot}<1.2\times10^{12}$;][]{mwmass1,mwmass2} 
representing intermediate cases.
We argued that most
of the observed differences between colors involving the $FUV$ band
are explainable invoking different Helium contents.  This would lead
us to speculatively think that galaxies with larger masses may have, on
average, more He-rich populations. In that case, He abundance differences could be 
a by-product of chemical evolution differences, in some way connected 
to the mass of the host galaxy. This could be also connected with the 
formation and dynamical history of
clusters in galaxies with different masses, as suggested by Valcarce \& Catelan (2011). 
In particular they argue
that clusters hosted by more massive galaxies are more likely to undergo a more complex history of
star formation thus having a larger spread in stellar populations properties.

\section{Summary}
As part of a project aimed at studying the properties of hot stellar
populations in the Milky Way GCs (see Paper I), we have presented UV
integrated colors obtained with GALEX for 44 clusters spanning a wide
range of metallicities ($-2.5<\feh<-0.4$), HB morphologies, structural
and dynamical parameters. This represents the largest homogeneous
catalog of UV photometry ever built for GGCs.

We compared the behavior of UV colors with several parameters characterizing the 
morphology of the HB.
As expected, there are general correlations, in
particular between $(FUV-V)_0$ and the $(B2-R)/(B+V+R)$ parameter defined by 
Buonanno et al. (1993; 1997) and the HB temperature extension
\teff\ defined by Rec\'\i o Blanco et al. (2005). There is also a
significant correlation with the $\Delta (V-I)$ parameter (Dotter et
al. 2010), but, as expected, this parameter is insensitive to
HBs with extreme blue extensions.

In all color combinations, the bluest clusters are those in the
intermediate metallicity regime ($-1.5<\feh<-1$). This is in agreement
with DOR95 and with the L$_{\rm t}$
parameters measured by Fusi Pecci et al. (1993).  In the $(NUV-V)_0$-$\feh$ plane,
clusters more metal rich than $\feh\sim-1.5$
show a clear and significant linear correlation with cluster becoming redder as
metallicity increases, while there is an
opposite trend in the metal-poor regime. The combinations of colors
involving the $FUV$ appear more scattered, but a reasonable and similar correlation 
with metallicity at ($2.5\sigma$ level) has been also found in these cases.
All the clusters suspected to be connected with the
Sagittarius dwarf spheroidal (NGC~4590, NGC~5053, NGC~5466, Arp~2 and
Terzan~8) are typically $\sim 1.5$\,mag redder in $FUV$ colors 
than systems with similar iron content. 
No appreciable differences are found in $(NUV-V)_0$. 
Studies from different groups suggest that Sagittarius clusters and their 
galactic counterparts are coeval, while GGCs are on average more He-rich
than the Sagittarius sub-set. This would tentatively be interpreted as due to
a different environment in which they formed.

With the aim of showing how sensitive ages derived from UV colors
may be to assumptions about helium abundance, 
we compared our colors with evolutionary models of SSP by Lee et
al. (2002).  The color-metallicity distribution of GGCs can be
reproduced by assuming an average age of $\sim 12$ Gyr, with a spread
of about $\pm 2\,\gyr$. Alternatively, in the framework in which some
GCs have experienced self-enrichment from material ejected from AGBs
or fast-rotating massive stars (Ventura \& D'Antona 2008; Decressin et
al. 2007), the color spread is consistent with different He
content. In particular, we show that an overabundance of helium
($\Delta Y(R') \sim 0.05 $ ) can mimic an age difference $\Delta t
\sim 2\,\gyr$.

The UV colors of GGCs are consistent with those obtained by GALEX for
M31 clusters (Rey et al. 2007; Kang et al. 2011), at least in the intermediate/low
metallicity regime. At $\feh>-1$, M31 GCs are systematically bluer by
1--2 mag, behaving like massive MW clusters with bimodal HBs, such as
NGC~6388 and NGC~6441. As already noticed by Sohn et al. (2006), M87
GCs are on average bluer than GGCs.  This might be the signature
of different chemical abundances impressed on the "integrated" properties
of GCs. In particular we speculate that He abundance may be correlated with the mass of the host
galaxy, being higher in GCs belonging to higher mass galaxies.\\


{\it The authors dedicate this paper to the memory of co-author Bob Rood, a pioneer in the theory of the
evolution of low mass stars, and a friend, who sadly passed away on 2 November 2011.}\\
We thank the anonymous referee for the useful comments and suggestions.
The authors warmly thank M. Bellazzini and F. Fusi Pecci for useful discussions 
that improved the presentation of the results. We are grateful to M. Catelan 
for providing us informations about HB parameters.
This work is based on observations made with the NASA Galaxy Evolution Explorer.
GALEX is operated for NASA by the California Institute of Technology
under NASA contract NAS5- 98034.  This research is part of the project
COSMIC-LAB funded by the European Research Council (under contract
ERC-2010-AdG-267675).  E.D. thanks the hospitality and support from
Gemini Observatory where most of this work was developed, during two
extended visits in 2009 and 2010, and a shorter visit in 2011.
R.P.S. acknowledges funding by GALEX grants $\#$ NNG05GE50G and
NNX08AW42G and support from Gemini Observatory, which is operated by
the Association of Universities for Research in Astronomy, Inc., on
behalf of the international Gemini partnership of Argentina,
Australia, Brazil, Canada, Chile, the United Kingdom, and the United
States of America.  S.T.S. acknowledges support for this work from the
GALEX Guest Investigator Program under NASA grant NNX07AP07G.

\newpage

\newpage

\newpage

\newpage


\newpage

\newpage


\newpage

\newpage
\begin{figure}[!hp]
\begin{center}
\includegraphics[scale=0.8]{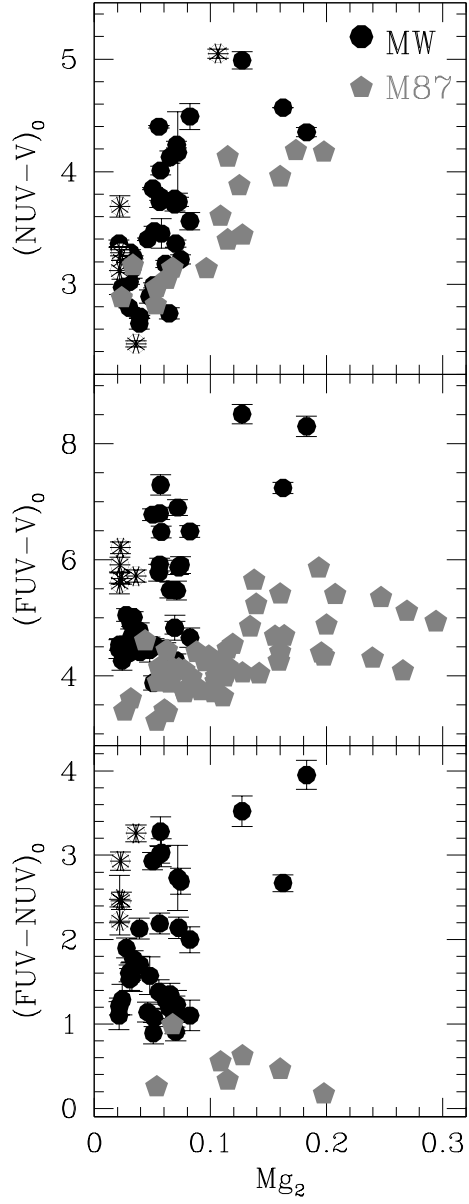}
\caption{UV colors of GGCs (black dots and asterisks) compared to those of M87
(grey pentagons) observed by Sohn
  et al.  (2006), as a function of the $Mg_2$ metallicity index.}
\label{m87}
\end{center}
\end{figure}

\begin{deluxetable}{cccccccc}
\scriptsize
\tablewidth{16.5cm}
\startdata \\
\hline \hline
CLUSTER    &   \feh\  & $E(B-V)$  &  $V_{t,0}$    &  $(FUV-NUV)_0$    &    $(FUV-V)_0$    &   $(NUV-V)_0$  \\
\hline
NGC 104   & -0.76     & 0.04  &   4.09    &    3.52   &  8.51  &   4.99 \\
NGC 1261  & -1.27     & 0.01  &   8.63    &    2.69   &  5.91  &   3.22 \\
NGC 1851  & -1.18     & 0.02  &   7.23    &    2.00   &  6.49  &   4.49 \\
NGC 1904  & -1.58     & 0.01  &   8.16    &    0.89   &  3.88  &   2.99 \\
NGC 2298  & -1.96     & 0.14  &   8.89    &    1.62   &  4.91  &   3.28 \\
NGC 2419  & -2.20    & 0.08  &   10.05   &    1.29   &  4.26  &   2.97 \\
NGC 288   & -1.32     & 0.03  &   8.13    &    0.91   &  4.27  &   3.36 \\
NGC 2808  & -1.18     & 0.22  &   5.69    &    1.10   &  4.66  &   3.56 \\
NGC 362   & -1.30     & 0.05  &   6.58    &    2.73   &  6.90  &   4.17 \\
NGC 4147  &  -1.78    & 0.02  &   10.74   &    1.70   &  4.41  &   2.71 \\
NGC 4590  &  -2.27    & 0.05  &    7.96   &    2.93   &  6.21  &   3.28 \\
NGC 5024  &  -2.06    & 0.02  &    7.79   &    1.90   &  5.05  &   3.15 \\
NGC 5053  &  -2.30    & 0.01  &    9.96   &    2.22   &  5.91  &   3.69 \\
NGC 5272  &  -1.50    & 0.01  &    6.39   &    2.19   &  5.92  &   3.73 \\
NGC 5466  &  -2.31    & 0.00  &    9.70   &    2.48   &  5.60  &   3.12 \\
NGC 5897  &  -1.90    & 0.09  &    8.52   &    1.77   &  5.01  &   3.24 \\
NGC 5904  &  -1.33    & 0.03  &    5.95   &    1.11   &  4.83  &   3.71 \\
NGC 5986  & -1.63     & 0.28  &   6.92    &    1.57   &  4.45  &   2.89 \\
NGC 6101  & -1.98     & 0.05&  10.08    &    1.53   &  4.56  &   3.02 \\
NGC 6218  & -1.33     & 0.19  &   6.07    &    1.07   &  4.83  &   3.76 \\
NGC 6229  & -1.43     & 0.01  &   9.86    &    1.29   &  4.47  &   3.18 \\
NGC 6235  & -1.38     & 0.31  &   7.20    &    1.35   &  5.48  &   4.13 \\
NGC 6254  & -1.57     & 0.28  &   4.98    &    1.07   &  4.53  &   3.47 \\
NGC 6273  & -1.76     & 0.38  &   5.57    &     ---   &  4.49  &	--- \\
NGC 6284  & -1.31     & 0.28  &   7.43    &    1.23   &  5.47  &   4.24 \\
NGC 6341  &  -2.35    & 0.02  &    6.52   &    1.10   &  4.46  &   3.36 \\
NGC 6342  &  -0.49    & 0.46  &   10.01   &    2.67   &  7.24  &   4.57 \\
NGC 6356  &  -0.35    & 0.28  &    7.42   &    3.95   &  8.30  &   4.35 \\
NGC 6397  &  -1.99    & 0.18 &    5.17   &    1.60   &  4.39  &   2.79 \\
NGC 6402  &  -1.39    & 0.60  &    5.73   &    1.19   &  3.93  &   2.74 \\
NGC 6535  &  -1.79    & 0.34  &    9.85   &    2.13   &  4.78  &   2.65    \\
NGC 6584  &  -1.50    & 0.10   &    8.17   &    3.28   &  7.29  &   4.01    \\
NGC 6809  &  -1.93    & 0.08   &    6.49   &    1.56   &  4.69  &   3.13    \\
NGC 6864  &  -1.29    & 0.16  &    8.26   &    2.14   &  5.87  &   3.73    \\
NGC 6981  &  -1.48    & 0.05   &    8.96   &    3.03   &  6.48  &   3.45    \\
NGC 7006  &  -1.46    & 0.05   &   10.46   &    2.93   &  6.78  &   3.85    \\
NGC 7089  &  -1.66    & 0.06  &    6.25   &    1.14   &  4.54  &   3.40    \\
NGC 7099  &  -2.33    & 0.03  &    7.10   &    1.21   &  4.54  &   3.33    \\
\hline
\hline
\enddata
\end{deluxetable}

\setcounter{table}{0}
\begin{deluxetable}{ccccccc}
\scriptsize
\tablewidth{16.5cm}
\tablecomments{Integrated reddening corrected UV colors and integrated V magnitudes 
obtained by fitting the surface brightness profiles (see Section 2.1 for more details). 
For reddening correction we used the following coefficients: $R_{FUV}=8.2$, $R_{NUV}=9.2$ 
and $R_V=3.1$ (Cardelli et al. 1989).
The adopted $E(B-V)$ values are from H10, while \feh\ value are from Carretta et al. (2009b).
For the three clusters with large discrepancies between SBP and AP magnitudes we report also 
AP colors. For 47~Tuc AP colors are $(FUV-NUV)_0=2.72$, $(FUV-V)_0=7.49$, $(NUV-V)_0=4.77$.
For NGC~1851 $(FUV-NUV)_0=1.14$, $(FUV-V)_0=4.95$, $(NUV-V)_0=3.81$ and for NGC~6864 
$(FUV-NUV)_0=1.60$, $(FUV-V)_0=4.73$, $(NUV-V)_0=3.13$.}
\startdata \\
\hline \hline
CLUSTER    &   \feh\    & $E(B-V)$  &  $V_{t,0}$    &  $(FUV-NUV)_0$    &    $(FUV-V)_0$    &   $(NUV-V)_0$  \\
\hline
Pal 11    &  -0.45    & 0.35  &    7.54   &    ---   &     ---    &     ---   \\
Pal 12    &  -0.81    & 0.02  &   11.89   &    ---   &     ---    &   5.05    \\
NGC 7492  &  -1.69    & 0.00  &   10.48   &   1.38  &    5.78     &   4.40  \\
IC  4499  &  -1.62    & 0.23  &    8.56   &   3.01  &    6.80     &   3.78  \\
Terzan 8  &  -2.22    & 0.12  &   11.54   &   2.46  &    5.68     &   3.21  \\
Arp 2	  &  -1.74    & 0.10  &   12.41   &   3.26  &    5.72     &   2.47  \\
\hline	   
\hline     
\enddata  
\label{tab:masses}
\end{deluxetable}

\end{document}